\documentclass[11pt]{article}

\usepackage{amsmath,amssymb,amsthm,graphics,amsfonts,graphicx,float,subfigure}
\usepackage{lineno}
\newtheorem{theorem}{Theorem}[section]

\begin{document}

\begin{center}  {\Large \bf Analysis of a mosquito--borne epidemic model with vector stages and saturating forces of infection}
\end{center}

\smallskip

\smallskip
\begin{center}
{\small \textsc{Eric \'Avila--Vales}\footnote{Corresponding author. email: avila@uady.mx},
\textsc{Bruno Buonomo}\footnote{email: buonomo@unina.it},
\textsc{No\'e Chan--Ch\i}\footnote{email: noe.chan@uady.mx} }
\end{center}
\begin{center} {\small \sl $^{1,3}$ Facultad de Matem\'aticas, Universidad Aut\'onoma de Yucat\'an, \\ Anillo Perif\'erico Norte, Tablaje 13615, C.P. 97119, M\'erida, Mexico}\\ {\small \sl $^{2}$ Department of Mathematics and Applications,
University of Naples Federico II,\\  via Cintia, I-80126 Naples,
Italy}
\end{center}

\bigskip

{\small \centerline{\bf Abstract}
\begin{quote}
We study a mosquito--borne epidemic model where the vector population is distinct in aquatic and adult stages and a saturating effect of disease transmission is assumed to occur when the number of infectious (humans and mosquitoes) becomes large enough. Several techniques, including center manifold analysis and sensitivity analysis, have been used to reveal relevant features of the model dynamics. We determine the existence of stability--instability thresholds and the individual role played in such thresholds by the model parameters. 
\end{quote}}

\medskip

\noindent {\bf Keywords}: Mosquito--borne disease; endemic equilibrium; stability analysis; sensitivity analysis

\medskip

\noindent {\bf AMS classification}: 92D30, 34D20

\bigskip

\section{Introduction}
\label{intro}

Mosquito--borne diseases are caused by pathogens transmitted among hosts by mosquitoes (the \emph{vector}). Responsible of serious illness are mosquitoes of genus \emph{Anopheles} (malaria, filariasis), \emph{Aedes}  (yellow fever, dengue, chikungunya), and \emph{Culex}  (West Nile, Japanese encephalitis, filariasis)  \cite{to}. Mosquitoes are distributed globally in the world, but invasive species colonize new habitats affecting the ecology and economy of those areas \cite{beetal}. As relevant example, the continuing spread of the mosquito \emph{Aedes albopictus} in Europe is a big health concern due to the potential risk of new outbreaks of exotic diseases that this species can transmit \cite{neetal,poetal}. A similar problem concerns with the population expansion of mosquito \emph{Aedes aegypti} in Brasil \cite{glgo}.  For this reason, many studies are focused on  prediction of short and long term invasion by vectors, their impact on the invaded areas and invasion control. Mathematical models describing the dynamics of the competent vectors are among the main tools to provide estimates of the transmission potential of viruses and to assess the efficacy of the measures undertaken by public health authorities to control the epidemic spread \cite{adetal,luetal,maya,medebr,moaaca,poetal,romoto,taetal}.

In 2005 a mosquito--borne epidemic model has been proposed by N.A. Maidana, H.M. Yang and coworkers \cite{taetal}. Their model is given by five nonlinear ordinary differential equations (ODEs) and couples the dynamics of humans and mosquitoes. The former group is divided into two compartments (susceptibles and infectious) while the latter is divided in three compartments, since the vector's aquatic stage is explicitly considered, together with adult stage (susceptibles and infectious). An important feature of this five equation model is that it has been the basis for studying a real case: the \emph{Aedes aegypti} dispersal dynamics in the state of Sao Paulo, Brasil, and the consequent dissemination of dengue virus \cite{maya,taetal}.

A fundamental issue of epidemic modeling is the description of disease transmission. A key role in modeling this mechanism is played by the \emph{force of infection} (FoI), which is defined as the per capita rate at which susceptible individuals contract the infection \cite{kero}. In many cases, the FoI is assumed to be proportional to the size of infectious compartment (see e.g. \cite{anma,he,kero}). In particular, such an assumption is made in the abovementioned model \cite{maya, taetal}. However, since the seventies, V. Capasso and his coworkers stressed the importance to consider nonlinear FoI \cite{cap, cagrse, case}. Since then, various nonlinear forms of FoI have been proposed by many authors (see e.g. the brief surveys contained in \cite{do} and \cite{xiru}).

As a matter of fact, as it has been underlined in \cite{jiwaxi}, the details of transmission of infectious diseases are generally unknown and may depend on several factors. In particular, for vector--host epidemics, there are several biological mechanisms which may result in nonlinearities in the transmission rates of parasites \cite{cali}. For this reason, several authors have recently proposed nonlinear forces of infection for vector--host epidemics, for one or both the transmissions (from vector to host and viceversa) \cite{bbcruz2,cagu,cali,esma,ozetal}.

Motivated by the above discussion, in this paper we extend the model proposed in \cite{maya,taetal} by assuming that the forces of infection are nonlinear. In particular, we assume that there is a saturating effect of diseases transmissions when the number of infectious (humans and mosquitoes) becomes large enough \cite{cagu,cali,ozetal}. Therefore, we adopt Holling II functional responses to represent the forces of infection, so that the incidence rate generalizes the simple mass action law \cite{anma,he,kero}. We perform a qualitative analysis which enhances the one performed in \cite{taetal}, where the analysis of the ODEs model is only sketched. 

From a mathematical point of view, considering both the two aspects, two--stages vector population and nonlinear forces of infection, leads to several open problems. It can be seen that, due to model complexity, even the complete analysis of local stability of endemic states is an open task. Several techniques, including centre manifold analysis \cite{guho} and sensitivity analysis \cite{maetal}, have been here adopted to reveal relevant features of the model dynamics like the existence of stability--instability thresholds and the individual role played in such thresholds by the model parameters. 

The plan of the paper is organized as follows. In Section 2, we formulate the model. Existence and linear stability analysis of the disease free equilibria are also carried out. In Section 3, we study the endemic states through stability and bifurcation analyses. We perform local and global sensitivity analyses in Section 4. Finally we draw our conclusions in Section 5.

\section{The model, basic properties and disease--free states}
\label{model}

We consider two interacting populations, mosquito and humans. The mosquito population is divided into two subpopulations, the winged form and aquatic form. Regarding the winged mosquitoes, susceptible and infectious are designated by $M_S$ and  $M_I$, respectively. Aquatic subpopulation is denoted by $A$. Susceptible and infectious humans are denoted by $H$ and $I$, respectively. The balance equations lead to the following system of nonlinear ordinary differential equations:
\begin{equation}\label{ec1}
\begin{array}{ll}
\dot M_S=  {\displaystyle\frac{\gamma}{k}} A \left(1-M_S-M_I\right)-\mu _1 M_S- \beta _1 g_1(I) M_S \\
\dot M_I= \beta _1 g_1(I) M_S  - \mu_1 M_I\\
 \dot A = k\left(1-A\right)\left(M_S+M_I\right)-\mu _2A-\gamma A \\
 \dot H = \mu _H-\mu _H H-\beta _2 g_2(M_I) H \\
 \dot I = \beta _2 g_2(M_I) H  - \left(\sigma+\mu_H\right) I,
\end{array}
\end{equation}
where the upper dot denotes the time derivative. All the parameters involved in the model are positive constants. The parameter $\beta _1$ is the rate at which susceptible mosquitoes are infected when they bite infectious humans; $\beta_2$ is the rate at which susceptible humans are infected when they are bitten by infectious mosquitoes. These transmission coefficients are given by $\beta_1=b \beta_V$, where $b$ is the average biting rate and $\beta_V$ is the average transmission probability from human to vector, and $\beta_2=b \beta_H$, where $\beta_H$ is the average transmission probability from vector to human.\\
The meaning of all the parameters of (\ref{ec1}) are summarized in Table \ref{param}.

\begin{table}
\centering
\begin{tabular}{llll} 
\hline\noalign{\smallskip}
Parameter & Description & Baseline value\\
\noalign{\smallskip}\hline\noalign{\smallskip}
$k$ & ratio between carrying capacities  &   \\
    & of winged and aquatic form         & 0.25     \\
$\beta_1$ & rate of effective contact between &   \\
& uninfected mosquitoes and infected humans        & 0.3265    \\
$\beta_2$ & rate of effective contact between &     \\
& uninfected humans and infected mosquitoes        & 0.0411   \\
$\gamma$ & Inverse of period of time in aquatic form & 0.0125 \\
$\mu_1$ & Inverse of survival time in winged phase & 0.025 \\
$\mu_2$ & Inverse of survival time in aquatic phase & 0.1096 \\
$\mu_H$ & Inverse of life expectancy in humans & 0.00003 & \\
$\sigma$ & Inverse of human infectious period & 0.1096 & \\
$\alpha_1$ & Holling II parameter for infectious humans & 0.3 \\
 $\alpha_2$ & Holling II parameter for infectious mosquitoes & 0.3 \\
\noalign{\smallskip}\hline
\end{tabular}
\caption{\label{param} \small Description and baseline values of parameters in system (\ref{model}). All the values are non dimensional. The values of $\alpha_1$ and $\alpha_2$ are guessed. All the other values are the non dimensional version of the values in  Tables 2 and 3 in \cite{maya}, after a rescaling with respect to the oviposition rate $r=1.25$ (see \cite{maya} for details and specific references for the chosen values.}
\end{table}

As mentioned and motivated in the previous section, we assume that there is a saturating effect of diseases transmissions when the number of infectious (humans and mosquitoes) becomes large enough \cite{cagu,cali,ozetal}.  Therefore, we adopt Holling II functional responses to represent the forces of infection,
\begin{equation}\label{FoI}
g_1(I)=\frac{I}{1+\alpha _1I},\;\;\;\;\;\;\;\;g_2(M_I)=\frac{M_I}{1+\alpha _2 M_I}.
\end{equation}
It follows that the incidence rate generalizes the mass action law \cite{anma,he,kero}. When $\alpha_1=\alpha_2=0$, model (\ref{ec1}) reduces to the one proposed in \cite{maya,taetal}. \\
Finally, initial conditions 
\begin{equation} \label{ic}
M_S(0)>0, \;\; M_I(0)>0, \;\; A(0)>0, \;\; H(0)>0 \;\; I(0)>0,
\end{equation}
are appended to model (\ref{ec1}).

It is easy to check that the feasible region for (\ref{ec1}) is the positive orthant of ${\bf R}^5$, and that the closed 
set
$$
\Omega=\left\{\left(M_S,\;M_I,\; A,\; H,\;I\right)\in{\bf R}^5_+: M_S+M_I\leq 1, \; A\leq 1, \; H+I \leq 1 \right\}
$$
is positively invariant and attracting with respect to the solutions of model (\ref{ec1}) and, as a consequence, the orbits
of (\ref{ec1}) are bounded, provided that the initial conditions are given by (\ref{ic}).

As shown in the appendix \ref{DFE}, model  (\ref{ec1}) admits two disease--free equilibria. The first one is
\begin{equation}
\label{E0}E_0\equiv(M_S^0,M_I^0,A^0,H^0,I^0)=(0,0,0,1,0),
\end{equation}
which corresponds to the presence of only human population, without mosquitoes. This is a trivial equilibrium (\emph{mosquito--free} and \emph{disease--free}). The second disease--free equilibrium corresponds to coexistence of humans and mosquitoes, without infection. It is given by
\begin{equation}
\label{E1}
E_1\equiv(M_S^1,M_I^1,A^1,H^1,I^1)=\left(m^*,0,a^*,1,0\right),
\end{equation}
where 
\begin{equation} \label{ec2}
m^*= \dfrac{\gamma (1-Q_{0}^{-1})}{\gamma +k\mu_1}, \quad a^*=\dfrac{k(1-Q_0^{-1})}{k+\mu_2+\gamma},
\end{equation}
and
\begin{equation} \label{Q0}
Q_0=\dfrac{\gamma}{\mu _1(\gamma+\mu_2)}.
\end{equation}
It follows that $E_1$ is biologically feasible only if $Q_0>1$.

Following the procedure and the notation in \cite{vawa02}, we may obtain the basic reproduction number $R_0$, which may be obtained as the dominant eigenvalue (more precisely the spectral radius) of the \emph{next--generation matrix} \cite{dihe,vawa02}. Observe that model \eqref{ec1} has two infected populations, namely $M_I$ and $I$. It follows that the matrices $F$ and $V$ defined in \cite{vawa02}, which take into account of new infection terms and remaining transfer terms, respectively, are given by
\[
F= \begin{pmatrix} 0 & \beta _1m^* \\ \beta _2 & 0 \end{pmatrix},\;\;\;\;\;\;\; V=\begin{pmatrix} \mu _1 & 0 \\ 0 & \sigma +\mu _H 
\end{pmatrix}
\]
The next--generation matrix is the matrix 
\[
FV^{-1}= \begin{pmatrix} 0 & \frac{\beta _1 m}{\sigma +\mu _H} \\ \frac{\beta _2}{\mu _1} & 0 \end{pmatrix}
\]
and the dominant eigenvalue of $FV^{-1}$ is
\[
R_0 = \sqrt{\frac{\beta _1 \beta _2m^*}{\mu _1 (\sigma +\mu _H)}},
\]
or, in terms of $Q_0$ defined in \eqref{Q0},
\begin{equation}
\label{R0}
R_0 = \sqrt{\dfrac{\beta _1\beta _2\gamma(1-Q_0^{-1})}{\mu _1(\sigma +\mu_H)(\gamma +k\mu _1)}}.
\end{equation}
A direct consequence of the procedure given in \cite{dihe,vawa02} is the following:

\begin{theorem} \label{stabE1}
The disease--free equilibrium $E_1$, given by \eqref{E1}, is locally
asymptotically stable if $R_0 < 1$ and unstable if $R_0 > 1$, where $R_0$ is given by \eqref{R0}.
\end{theorem} 

By using standard linearisation procedure, it can be established that the threshold $Q_0=1$ is a threshold for the mosquitoes invasion, as stated by the following:

\begin{theorem}
\label{stabE0}
The trivial equilibrium $E_0$, given by \eqref{E0}, is locally asymptotically stable if $Q_0<1$ and unstable if $Q_0>1$.
\end{theorem}

The results of this section can be collected in the following sentence: If $Q_0<1$, then model \eqref{ec1} admits only the trivial  equilibrium $E_0$, given by  \eqref{E0}, which is locally asymptotically stable. If $Q_0>1$, then $E_0$ is unstable and the mosquitoes invasion takes place. In this case there exists the disease--free equilibrium $E_1$, given by  \eqref{E1}, which is locally asymptotically stable if $R_0<1$ and unstable if $R_0>1$. Therefore $R_0=1$ is a threshold for epidemic outbreak.

\section{Endemic equilibrium}

As \emph{endemic} we mean an equilibrium of system (\ref{ec1}) with all positive components. It is easy to check that model (\ref{ec1}) admits only one endemic equilibrium, $E^*=(M_S^\ast,M_I^\ast,A^\ast,H^\ast,I^\ast)$, where
\begin{equation}\label{endemic}\begin{aligned}
M_S^*&=m^*-M_I^*, \;\;\;\;\;\; M_I^*=\frac{\mu _H(1-H^*)}{\beta_2 H^* -\alpha _2\mu _H (1-H^*)}, \;\;\;\;\;\; A^*=a^*, \\
H^*&=1-\frac{\sigma +\mu _H}{\mu _H}I^* ,   \;\;\; I^*=\frac{\mu _1 \mu _H(R_0^2-1)}{[\beta _1m^* (\beta _2+\alpha _2\mu _H)+\mu _H (\beta _1+\alpha _1\mu _1)]},
\end{aligned}\end{equation}
and $m^*$ and $a^*$ are given by \eqref{ec2}.

The characteristic polynomial of the Jacobian matrix corresponding to \eqref{ec1} evaluated at $E_1$ is a fifth-degree polynomial. Using suitable mathematical software packages, we get
\begin{equation}
\label{caratt}
(\lambda ^2 +b_1\lambda +b_0)(\lambda ^3+a_2\lambda ^2+a_1\lambda+a_0)=0,
\end{equation}
where
\[ \begin{aligned}
b_0 & = (\gamma +k\mu _1)(M_I^*+M_S^*)+\frac{\gamma A^*}{k}(\gamma +\mu _2+k)+\gamma ( Q_0^{-1}-1),\\
b_1 & = \gamma +\mu _1+\mu _2+\frac{\gamma A^*}{k}+k(M_I^*+M_S^*),
\end{aligned}\]
and
\[ \begin{aligned}a_0 =& \mu _1\mu _H(\sigma +\mu _H)+\frac{\beta _2 M_I^*\mu _1}{\alpha _2M_I^*+1}(\sigma +\mu _H)+\frac{\beta _1I^*\mu _H}{\alpha _1 I^*+1}(\sigma +\mu _H)\\
&+\frac{\beta _1\beta _2M_I^*I^*}{(\alpha _1I^*+1)(\alpha _2M_I^*+1)}(\sigma +\mu _H)-\frac{\beta _1\beta _2\mu _HH^*M_S^*}{(\alpha _1I^*+1)^2(\alpha _2M_I^*+1)^2},\\
a_1 = & \mu _1\mu _H +(\sigma +\mu _H)(\mu _1+\mu _H)+\frac{\beta _2M_I^*}{\alpha _2M_I^*+1}(\sigma +\mu _1+\mu _H)+\frac{\beta _1I^*}{\alpha _1I^*+1}(\sigma +2\mu_H) \\
& \frac{\beta _1\beta _2 I^*M_I^*}{(\alpha _1I^*+1)(\alpha _2M_I^*+1)}-\frac{\beta _1\beta _2 H^*M_S^*}{(\alpha _1I^*+1)^2(\alpha _2M_I^*+1)^2}, \\
a_2 = & \sigma +\mu _1 + 2\mu _H+\frac{\beta _2 M _I^*}{\alpha _2M_I^*+1}+\frac{\beta _1I^*}{\alpha _1I^*+1}.
\end{aligned}\]
Taking into account that $A^*=a^*$, from (\ref{ec2}) it follows that $b_0 = (\gamma +k\mu _1)(M_I^*+M_S^*)$. Therefore, $b_0$ and $b_1$ are positive when $Q_0>1$ and, according to Routh--Hurwitz criterion, at least two eigenvalues of (\ref{caratt}) have negative real part.\\
Now observe that $a_2>0$ and that the second and fifth equations of \eqref{ec1} imply 
\begin{equation}
\label{MI}
M_I^*=\dfrac{\beta _1M_S^*I^*}{\mu _1(\alpha _1I^*+1)},
\end{equation} and 
\begin{equation}
\label{Istar}
I^*=\dfrac{\beta _2H^*M_I^*}{(\sigma +\mu _H)(\alpha _2M_I^*+1)}.
\end{equation} 
Substituting (\ref{MI}) and (\ref{Istar}) in the expression of $a_0$ and $a_1$ above, we get
\[ \begin{aligned}a_0 =& \mu _1\mu _H(\sigma +\mu _H)+\frac{\beta _2 M_I^*\mu _1}{\alpha _2M_I^*+1}(\sigma +\mu _H)+\frac{\beta _1I^*\mu _H}{\alpha _1 I^*+1}(\sigma +\mu _H)+\\
&+\frac{\beta _1\beta _2 H^*M_S^*}{(\alpha _1I^*+1)^2(\alpha _2M_I^*+1)^2}\left( \frac{\beta _1\beta _2 M_I^*I^*}{\mu _1(\sigma +\mu _H)} -1\right),\\
a_1 = & \mu _1\mu _H +(\sigma +\mu _H)(\mu _1+\mu _H)+\frac{\beta _2M_I^*}{\alpha _2M_I^*+1}(\sigma +\mu _1+\mu _H)+\frac{\beta _1I^*}{\alpha _1I^*+1}(\sigma +2\mu_H)+ \\
& +\frac{\beta _1\beta _2 H^*M_S^*}{(\alpha _1I^*+1)^2(\alpha _2M_I^*+1)^2}\left( \frac{\beta _1\beta _2 M_I^*I^*}{\mu _1(\sigma +\mu _H)} -1\right).
\end{aligned}\]
A sufficient condition ensuring that $a_0$ and $a_1$ are positive is that 
\begin{equation}
\label{SC}
\frac{\beta _1\beta _2 M_I^*I^*}{\mu _1(\sigma +\mu _H)}>1.
\end{equation}
As shown in the appendix \ref{Derivation}, this last inequality may be written in terms of the basic reproductive number, as
\begin{equation}
\label{condLAS}
R_0^2>1+\Delta,
\end{equation}
where the quantity $\Delta$ is given by
\begin{equation}
\label{Delta}
\Delta:=\frac{- K_2 + \sqrt{K_2^2+4K_0\beta_2}}{2K_0K_1},
\end{equation}
where
\begin{equation}
\label{K0}
K_0:=\frac{\beta_1\beta_2}{\mu_1},
\end{equation}
\begin{equation}
\label{K1}
K_1:=\frac{\mu_1\mu_H}{\beta_1m^*(\beta_2+\alpha_2\mu_H)+\mu_H(\beta_1+\alpha_1\mu_1)},
\end{equation}
and
\begin{equation}
\label{K2}
K_2:=\frac{(\beta_2+\alpha_2)(\sigma+\mu_H)}{\mu_H}.
\end{equation}

In other words, if condition \eqref{condLAS} holds, the Routh--Hurwitz criterion ensures that all the eigenvalues of (\ref{caratt}) have negative real part. This analysis can be summarized in the following:
\begin{theorem}
\label{stabEE}
The endemic equilibrium $E^*$, given by \eqref{endemic}, exists if $R_0>1$ and is locally asymptotically stable if $R_0>\sqrt{1+\Delta}$, where $\Delta$ is given by (\ref{Delta}).
\end{theorem}

The result stated in Theorem \ref{stabEE} gives only a sufficient condition for the local stability of the endemic equilibrium. It states that local stability is ensured for $R_0$ large enough. For this reason, we use a bifurcation theory approach to get an insight about the stability properties of the model near the criticality (at $E_1$ and $R_0=1$). In particular, we are interested to investigate if there is a stable endemic equilibrium bifurcating from the nonhyperbolic equilibrium $E_1$, and $E_1$ changes from being stable to unstable. This behaviour is called a 
\emph{forward} bifurcation \cite{ccso,duhucc,vawa02}. \\
To this aim, we study the centre manifold near the criticality (at $E_1$ and $R_0=1$) by using the approach developed in \cite{ccso,duhucc,vawa02}, which is based on the general centre manifold theory \cite{guho}. In short, this approach establishes that the normal form representing the dynamics of the system on the central manifold is given by 
$$\dot u = a u^2+ b \mu u,
$$
where,
\begin{equation}
\label{Coeff_a}
a=\frac{{\mathbf v}}{2} \cdot D_{{\mathbf x}{\mathbf x}}{\mathbf f}({\mathbf x}_0,0){\mathbf w}^2 \equiv \frac12 \displaystyle \sum_{k,i,j=1}^{n} v_{k} w_{i} w_{j} \displaystyle \frac{\partial^{2}f_{k}}{\partial x_{i} \partial x_{j}} ({\mathbf x}_0,0),
\end{equation}
and
\begin{equation}
\label{Coeff_b}
b= {\mathbf v} \cdot D_{{\mathbf x}\varphi}{\mathbf f}({\mathbf x}_0,0){\mathbf w} \equiv
\displaystyle \sum_{k,i=1}^{n} v_{k} w_{i} \displaystyle \frac{\partial^{2}f_{k}}{\partial x_{i} \partial \varphi} ({\mathbf x}_0,0).
\end{equation}

Note that in (\ref{Coeff_a}) and (\ref{Coeff_b}) $\varphi$ denotes a bifurcation parameter to be chosen, $f_k$'s denote the right hand side of system (\ref{ec1}), ${\mathbf x}$ denote the state vector, ${\mathbf x}_0$ the disease--free equilibrium $E_1$ and \textbf{v} and \textbf{w} denote, respectively, the left and right eigenvectors corresponding to the null eigenvalue of the Jacobian matrix of \eqref{ec1} evaluated at criticality (at ${\mathbf x}_0$ and $\varphi=0$).

In our case, let us choose $\beta_1$ as bifurcation parameter. Observe that $R_0=1$ is equivalent
to:
$$
\beta_1=\beta_1^*:=\frac{\mu_1\left(\sigma+\mu_H\right)\left(\gamma+k\mu_1\right)}{\beta_2\gamma \left(1-Q_0^{-1}\right)},
$$
so that the disease-free equilibrium $E_1$ is locally stable when $\beta_1<\beta_1^*$, and is unstable when $\beta_1>\beta_1^*$. Therefore, $\beta_1^*$ is a bifurcation value.

The direction of the bifurcation occurring at $\beta_1=\beta_1^*$ can be derived from the sign of coefficients (\ref{Coeff_a}) and (\ref{Coeff_b}). More precisely, if $a < 0$ and $b>0$, then at $\beta_1=\beta_1^*$ there is a forward bifurcation.\\
In our case we have the following (see the proof in appendix \ref{proofbif}):

\begin{theorem}
\label{Forward}
System (\ref{ec1}) exhibits a forward bifurcation at $E_1$ and $R_0=1$.
\end{theorem}

Putting together the results stated in Theorems \ref{stabEE} and \ref{Forward}, we know that the endemic equilibrium is locally stable near the criticality (i. e. for $R_0>1$ but $R_0-1<<1$) and that the stability is ensured for values of $R_0$ satisfying condition (\ref{condLAS}). It is useful to check that condition (\ref{condLAS}) could be, in principle, relaxed. To this aim we provide some numerical simulations.
 
\begin{figure}[t] \centering
\includegraphics[scale=0.5]{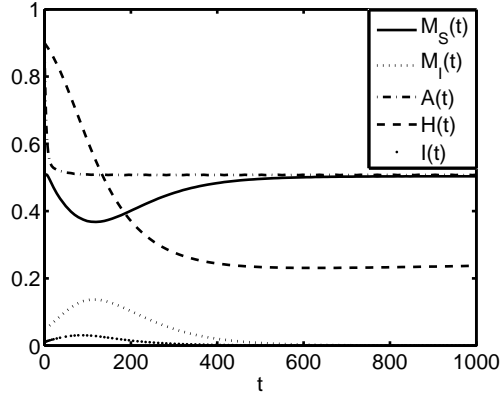}
\caption{\small Numerical solution of model \eqref{ec1}. The correspondence between lines and state variables is indicated in the label. The parameter values are chosen in way that condition (\ref{condLAS}) is verified, so that the system approaches the stable endemic equilibrium $E^*$. Note that here $E^*=(0.4984,0.0013,0.5056,0.3565,0.0020)$} \label{fig1}
\end{figure}

By using the parameter values in Table \ref{param}, we begin with the case in which condition (\ref{condLAS}) is verified.
In this case $Q_0=4.0910$ and $R_0=2.4579$ so that the condition of local stability is $R_0>1.2968$, which is verified. The dynamics of the model for this case is showed in Figure \ref{fig1}. 

\begin{figure}[t] \centering
\includegraphics[scale=0.5]{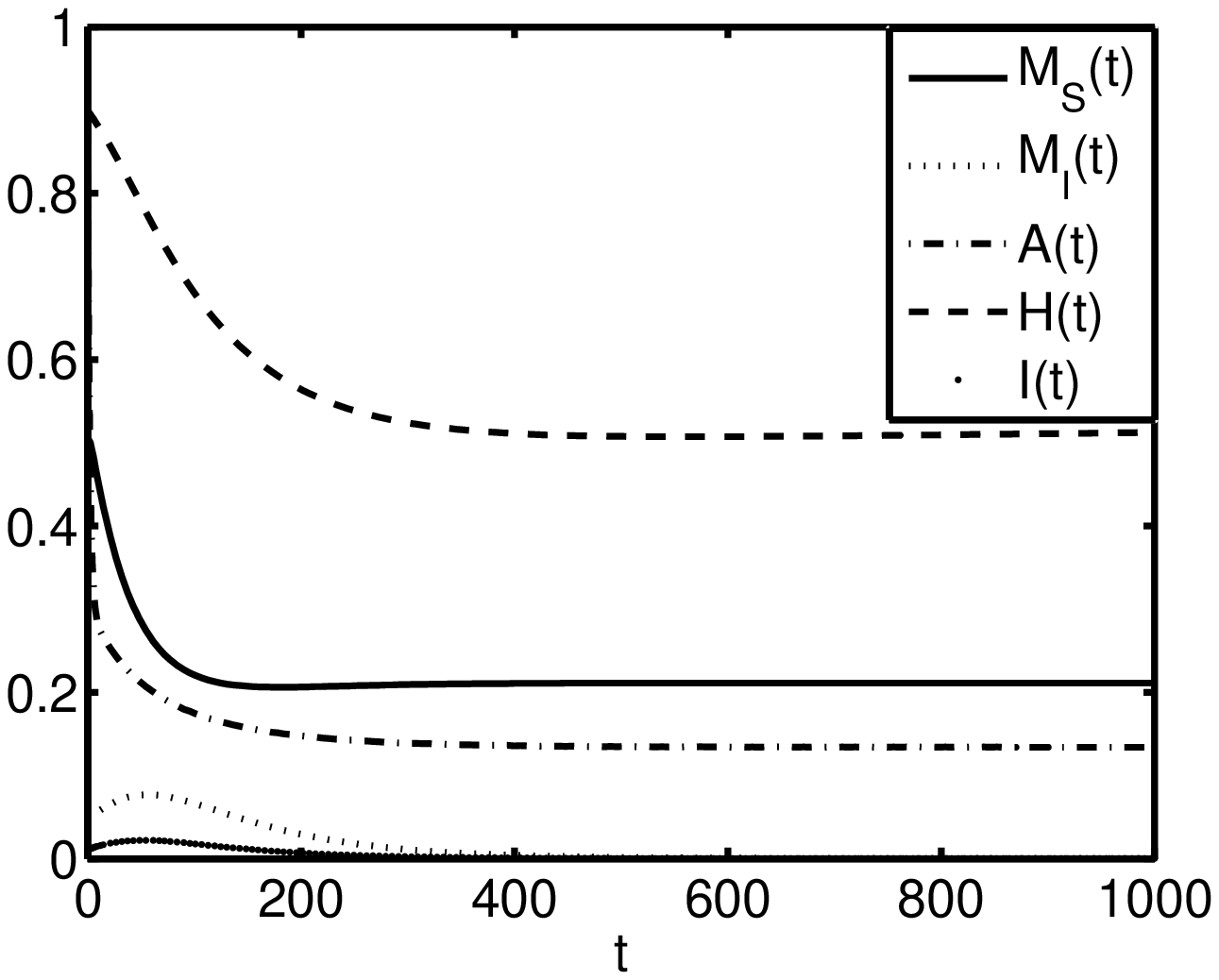} 
\caption{\small Numerical solution of model \eqref{ec1}. The correspondence between lines and state variables is indicated in the label. The parameter values are chosen in way that condition (\ref{condLAS}) is not verified. Nevertheless, the system approaches the stable endemic equilibrium $E^*=\left(0.2110,2.2\times 10^{-5},0.1338,0.9706,8\times 10^{-6}\right)$.} \label{fig2}
\end{figure}

As a second case we take again  the parameter values in Table \ref{param} with the exception of $\mu _2=0.3289$. In this case we have $Q_0=1.4635$, $R_0^2=1.0615$ and
$\Delta=0.1247$, where $\Delta$ is given by (\ref{Delta}). Therefore, the sufficient condition for local stability (\ref{condLAS}) is here not satisfied. However, as it can be seen in figure \ref{fig2}, the endemic equilibrium, which is given by $E^*=\left(0.2110,2.2\times 10^{-5},0.1338,0.9706,8\times 10^{-6}\right)$, is stable.

\section{Sensitivity analysis}\label{sensitivity}

In order to get an insight on the correct strategies to control the mosquito--borne epidemics described by model (\ref{ec1}), we 
perform a sensitivity analysis. We begin with a local sensitivity analysis and calculate the sensitivity indices of the basic reproduction number, in order to assess which parameter has the greatest influence on changes of $R_0$ and hence the greatest effect in determining whether the disease will be cleared in the population (see e.g. \cite{chitetal,chetal}).\\
To this aim, denote by $\Psi$ the generic parameter of model \eqref{ec1}. We calculate the \emph{normalised sensitivity index}, defined as the ratio of the relative change in $R_0$ to the relative change in the parameter $\Psi$:
$$
S_{\Psi}=\frac{\Psi}{R_0}\frac{\partial R_0}{\partial \Psi}.
$$
This index indicates how sensitive $R_0$ is to a change of parameter $\Psi$. Obviously, a positive (respectively negative) index indicates that an increase in the parameter value results in an increase (respectively decrease) in the $R_0$ value. We first note that:
\[ 
S_{\beta_1}=\frac{\beta_1}{R_0}\frac{\partial R_0}{\partial \beta_1}=
\frac{\beta_1}{R_0}\sqrt{\frac{\beta _2\gamma(1-Q_0^{-1})}{\mu _1(\sigma+\mu _H)(\gamma +k\mu _1)}}\;\frac{\partial\sqrt{\beta_1}}{\partial \beta_1}=\frac{1}{2},
\]
which means that $S_{\beta_1}$ does not depend on any parameter values. We also have:
$$
\begin{array}{c}
S_{\beta _2}=\frac{1}{2}, \;\;\; S_{\gamma}=-\frac{1}{2}\left( \frac{\gamma}{\gamma +k\mu _1}+\frac{\mu _2}{(\mu _2+\gamma)(1-Q_0)}\right) ,\\
S_{\mu _1}=\frac{1}{2}\left( \frac{\gamma +k\mu _1}{\mu _1(\mu _2+\gamma)(1-Q_0)}-1\right) ,\;\;\;
S_{\mu _2}=\frac{\mu _2}{2(\mu _2+\gamma)(1-Q_0)}, \\
S_{\sigma}=-\frac{\sigma }{2(\sigma +\mu _H)}, \;\;\;
S_{\mu _H}=-\frac{\mu _H}{2(\sigma +\mu _H)}, \;\;\;
S_{k}=-\frac{k\mu _1}{2(\gamma +k\mu _1)}.
\end{array}
$$

We evaluate the above sensitivity indices by using the parameter values in Table \ref{param}.

\begin{figure}[t]
\centering
\includegraphics[scale=0.50]{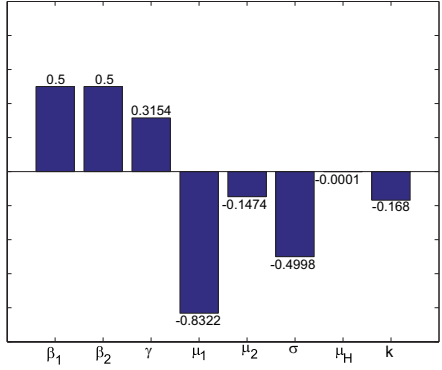}
\caption{Sensitivity indices of $R_0$ with respect to some chosen parameters.}
\label{figure1}
\end{figure}

As it can be seen in Figure \ref{figure1}, the basic reproductive number is most sensitive to the mortality rate of the mosquito winged form $\mu _1$, with $S_{\mu _1}=-0.8322$. This means that increasing $\mu _1$ by $10$\% will decrease $\mu _1$ by $8.32$\%. The other parameters with an important effect are the transmission parameters, $\beta _1$, $\beta _2$, and the recovery rate $\sigma$. Increasing or decreasing $\beta _1$ or $\beta _2$ by $10$\% will increase or decrease $R_0$ by $5$\%, and increasing $\sigma$ by $10$\% the value of $R_0$ will decrease by $4.9$\%

Local sensitivity analysis shows the effect of one parameter change while all others keep constant. In general, local sensitivity analysis is most informative when the model is linear or the range of possible values of the input factors is small. To obtain more accurate information, a global sensitivity analysis must be executed \cite{maetal}. Here, we used the ``sensitivity'' package of the software ``R'' \cite{Rpackage} to carry out the global sensitivity analysis of the reproduction number. The Latin Hypercube Sampling (LHS) Method was used to sample the input parameters using the parameter value ranges provided in Table \ref{Tab1}. Due to the absence of data on the distribution function,  a uniform distribution  was chosen for all parameters. The sets of input parameter values sampled using the LHS method, were used to run 10,000 simulations. We computed the Partial Rank Correlation Coefficients to estimate the correlation between $R_0$ and the parameters which define $R_0$. 

\begin{figure}[t]
\centering
\includegraphics[scale=0.50]{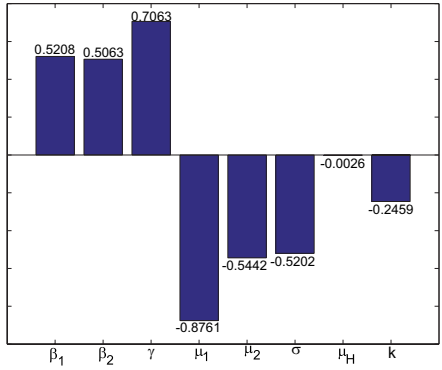}
\caption{Partial rank correlation coefficient showing the influence of parameter values variations on $R_0$. The parameter value ranges are given in Table \ref{Tab1}.}
\label{figure2}
\end{figure}

\begin{table}[t]
 \centering
\begin{tabular}{ll} \hline
Parameter & Range \\ \hline
$\beta_1$ & $[0.1381,0.3256]$ \\
$\beta_2$ & $[0.0131,0.0411]$ \\
$\gamma$ & $[0.0125,0.1315]$ \\
$\mu_1$ & $[0.0187,0.025]$ \\
$\mu _2$ & $[0.0229,0.3654]$ \\
$\sigma$ & $(0,0.11]$ \\
$\mu _H$ & $(0,0.00003]$ \\
$k$ & $[0.1,0.75]$ \\ \hline
\end{tabular}
\caption{\label{Tab1} Parameter value ranges used as input for the LHS method.}
\end{table}

The results, displayed in Figure \ref{figure2}, show that the parameters $\mu _1$ and $\gamma$ have the highest influence on the reproduction number $R_0$, while $\beta _1$, $\beta _2$, and $\sigma$ have similar influence that in the local sensitivity analysis. The partial rank correlation coefficient of the parameters $\gamma$, which represents the inverse of the period of time in mosquito aquatic form, and $\mu _1$, which represent the death rate of winged mosquitoes, shows a smaller influence than what was indicated by local sensitivity analysis. These discrepancies demonstrate the importance of the global sensitivity analysis in non-linear models.

\begin{figure}[t] \centering
\subfigure[$A^*$]{\includegraphics[scale=0.32]{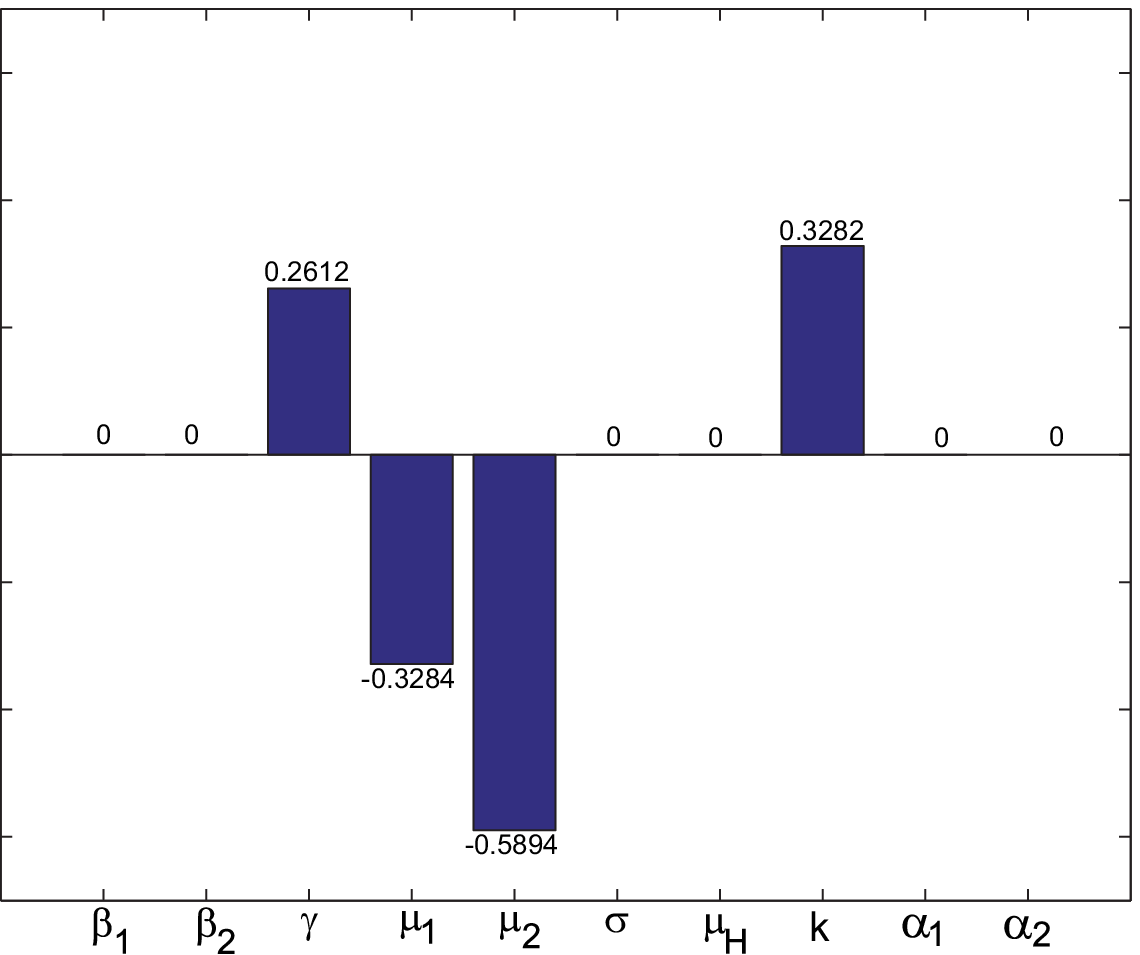}} \; \;
\subfigure[$M_S^*$]{\includegraphics[scale=0.32]{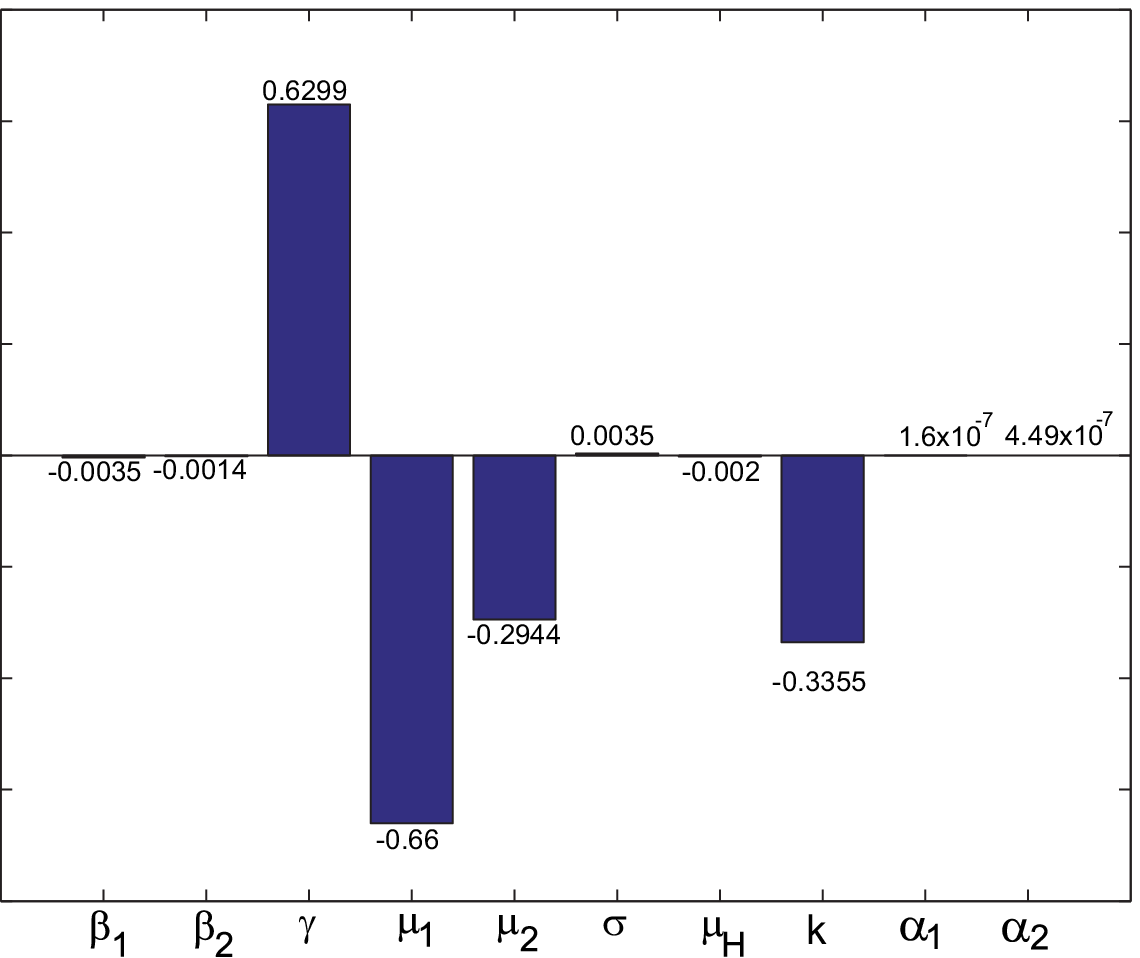}}\; \;
\subfigure[$M_I^*$]{\includegraphics[scale=0.32]{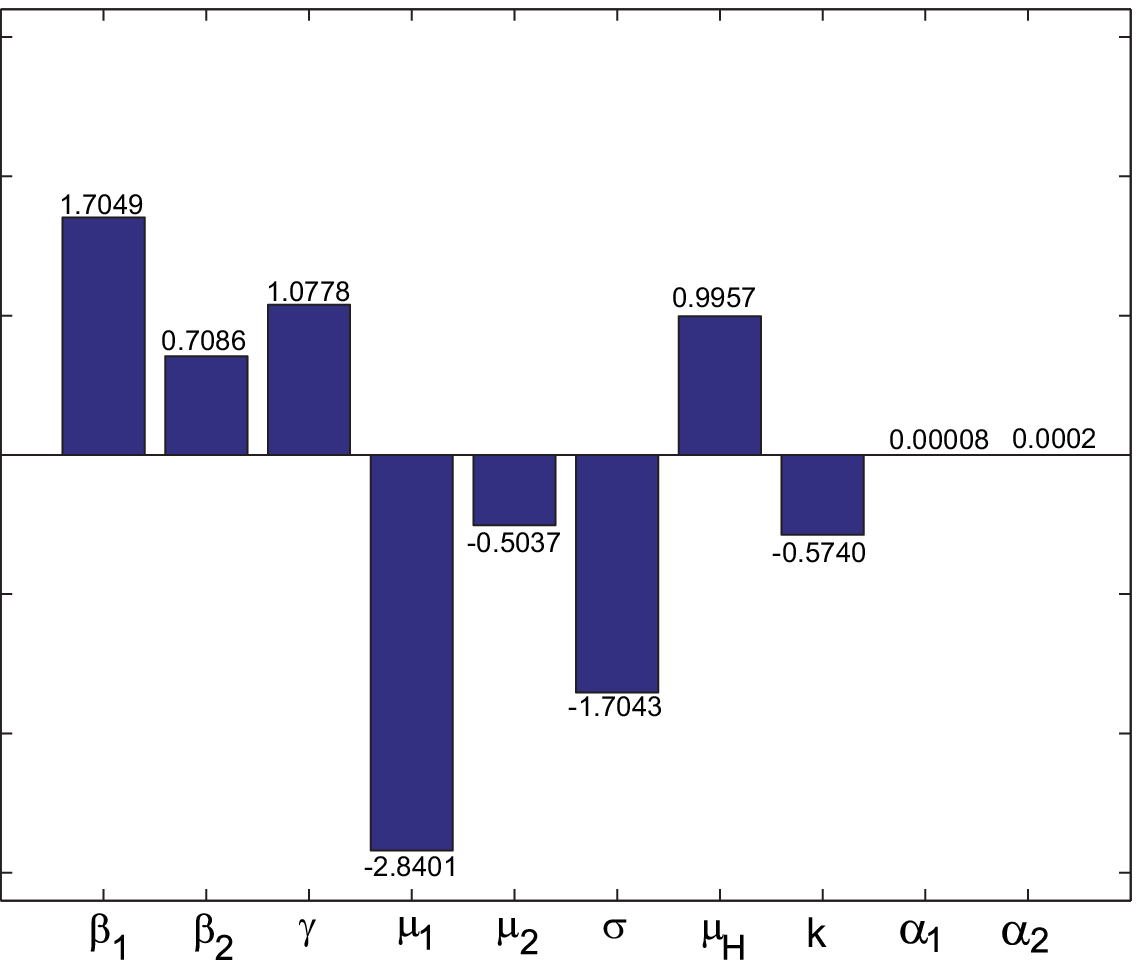}} \\
\subfigure[$H^*$]{\includegraphics[scale=0.32]{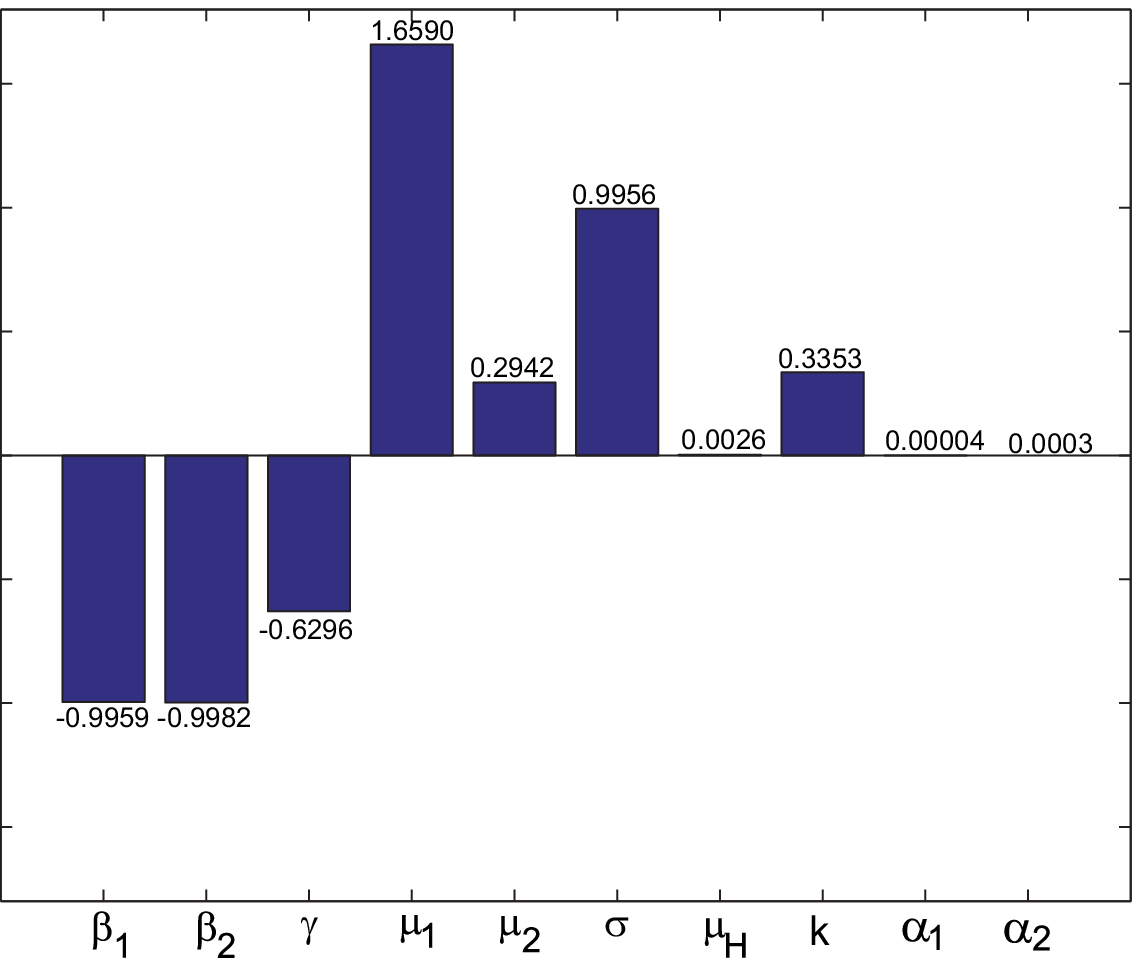}} \; \;
\subfigure[$I^*$]{\includegraphics[scale=0.32]{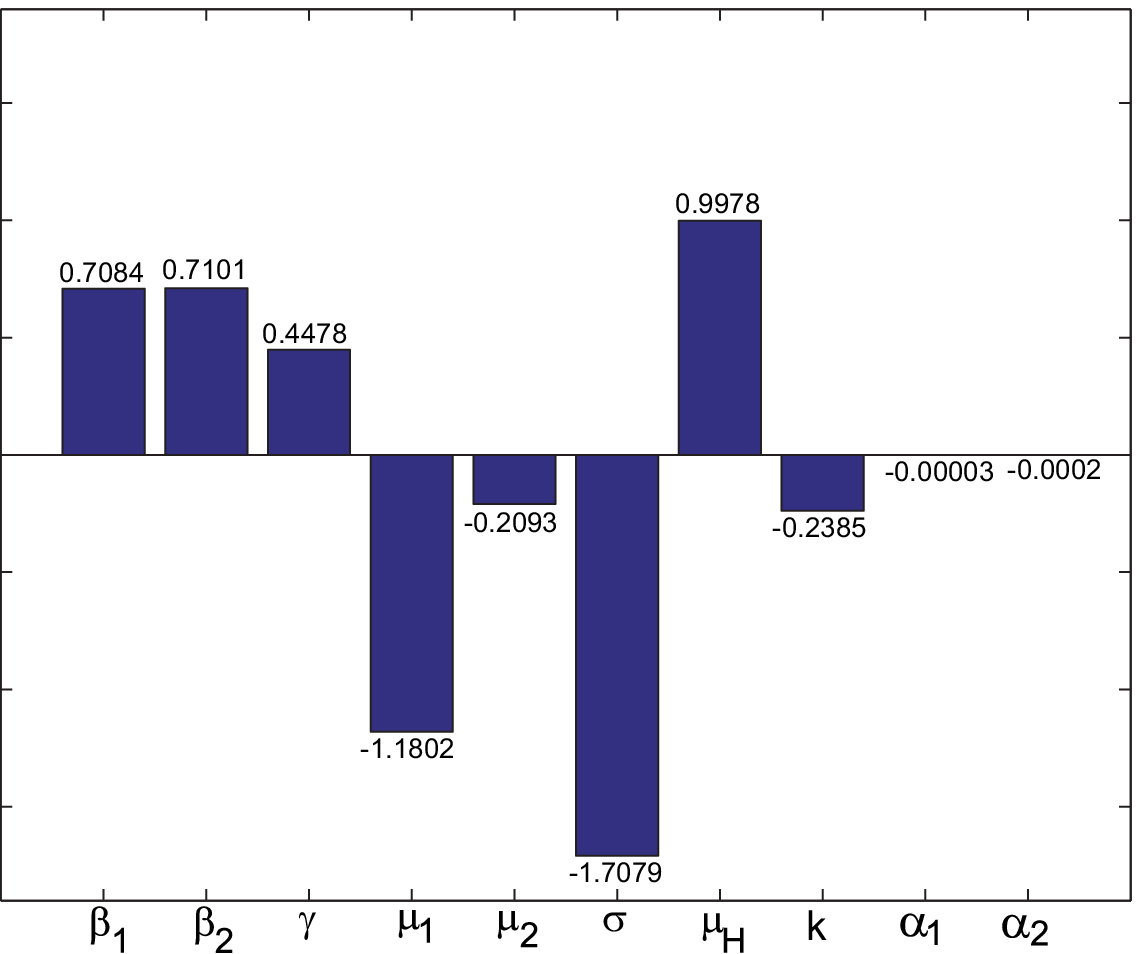}}
\caption{Sensitivity Index of the equilibria with respect to some chosen parameters} \label{indexee}
\end{figure}

We conclude this section by providing the sensitivity indices for the endemic equilibrium, still using the parameter values in Table \ref{param}. The results are showed in Figure \ref{indexee}. The positive number in the bar indicates a increase in the value of the equilibrium coordinate when the parameter increases and a negative indicates a decrease in the value when the parameter increases.
From Figure \ref{indexee} we can observe the following facts: for the value of $A^*$, showed in (a), the parameter with more influence is the death rate of the mosquito, $\mu _2$. By increasing this parameter by $10$\% will decrease the amount of mosquitoes in the aquatic form by $6$\%, note that the death rate of winged also decreases the value of $A^*$, by increasing $\gamma$ or $k$ will increase the amount of mosquitoes in aquatic phase by $2.6$\% and $3.2$\% respectively according to the increment of the parameter. From panel (b) we see that the parameter $\mu _1$ is the most influential for the value of the winged and susceptible mosquitoes, $M_S^*$, increasing the value of any parameter say by $10$\% the amount of winged mosquitoes will decrease by $6.6$\% in the case of $\mu _1$, on the other hand by increasing $\gamma$ the winged mosquitoes will increase by $6.2$\%, also increasing $\mu _2$ and $k$ the population of this class will decrease by $2.9$\% and $3.3$\% respectively. Similar results may be deduced from panels (c), (d), (e) of Figure \ref{indexee}.

\section{Conclusions}
\label{concl}

In this paper we have studied a mosquito--borne epidemic model, which is a generalization of the spatially homogeneous model considered in \cite{maya}. The model describes the interaction between humans and mosquitoes, the latter being divided in two stages, winged and aquatic. The forces of infections are represented by nonlinear functions of disease prevalence. In particular, we assume that there is a saturating effect of disease transmission when the number of infectious becomes large enough. From a mathematical point of view, considering both the two aspects, two--stages vector population and nonlinear forces of infection, enhances the model complexity and leads to several open problems, where even the complete analysis of local stability for the endemic state is an open task.  

We provide a qualitative analysis, based both on local stability analysis and bifurcation analysis, which shows that the threshold $R_0=1$ is a critical one. When $R_0$ is less than one, the disease cannot maintain itself in the population. Another parameter, namely $Q_0$, control the persistence or the elimination of the mosquitoes in the environment. On the other hand, when $R_0$ is greater than unity, the disease may persist in the population. In this case, by using a bifurcation approach, we are able to prove that a locally stable endemic equilibrium bifurcates from the disease--free equilibrium $E_1$ and that the stability property is maintained if the basic reproduction number is large enough (condition (\ref{condLAS})).

Our analysis demonstrates a certain robustness respect to the case of linear forces of infections treated in \cite{maya}, so that our results can be applied also to this last case, whose analysis is only sketched in \cite{maya}.

In order to explore effective control and prevention measures, we performed also a sensitivity analysis. The use of normalised sensitivity indices reveals that the basic reproductive number is most sensitive to the mortality rate of the mosquito winged form $\mu _1$. Changes of transmission parameters, $\beta_1$ and $\beta_2$, or of human recovery rate $\sigma$ have also important effects on $R_0$. This results has been mostly confirmed by global sensitivity analysis of $R_0$.

This suggest that intervention measures like mosquito adulticides (to increase $\mu_1$), use of  bed--net or other strategies that target the mosquito biting rate (to reduce $\beta_1$ and $\beta_2$) and treatment of infectious humans (to increase $\sigma$) are particularly effective to control the disease. This is coherent with other similar studies (see for example \cite{chitetal}).

The sensitivity analysis for the endemic equilibrium that we performed, shows that the parameter with more influence on the number of mosquitoes in aquatic stage is $\mu _2$, on the other hand for the winged type mosquitoes the parameter with more influence is $\mu _1$ which tell us that we need to increase the death rates to decrease the number of mosquitoes, while increasing the carrying capacity $k$ or $\gamma$ will increase the number of mosquitoes. In the case of infected mosquitoes again the death rate must increase to decrease the number of infected mosquitoes. For the human population, the parameters that increase the susceptible population are the death rate of mosquitoes and the recovery rate of the humans, we also have that these parameters decrease the infected  human population.

The previous analysis suggests that methods to decrease the biting rate and increase the death rate of winged mosquitoes are more effective than methods targeted to aquatic subpopulations (comprising eggs, larvae and pupae), as larvicides.

Finally, this model refers to specific locations where the mosquitoes have been already settled. It is well known that environment heterogeneity effect can be very important (see e. g. \cite{anca,dudu,maya}). In particular, diffusive and advective movements of mosquitoes and human movements may both increase the rate of mosquito--borne epidemic dissemination \cite{maya}. Therefore, the interplay between non--linear transmission and spatial heterogeneity may lead to more realistic model dynamics and improve the identification of correct control strategies to eradicate the disease or stop the vector invasion. We leave this issue for further studies.

\vspace{1cm}

\noindent {\bf Acknowledgements} The work of B. B. has been performed under the auspices of the Italian National Group for the Mathematical Physics (GNFM) of National Institute for Advanced Mathematics (INdAM). The work of E. A.--V. has been partially supported by SNI, under grant 15284. The authors are grateful to Marco and Ugo Avila for technical support.

\newpage

\appendix

\section{Existence of disease--free equilibria}
\label{DFE}

From \eqref{ec1} and \eqref{FoI} it follows that the steady states ${\overline E}=\left({\overline M}_S, \; {\overline M}_I, \; {\overline A}, \; {\overline H}, \; {\overline I}\right)$ are solutions of the algebraic system
\begin{equation}
 \frac{\gamma}{k}{\overline A}(1-{\overline M}_S+{\overline M}_I)-\mu _1 {\overline M}_S-\frac{\beta _1 {\overline M}_S {\overline I}}{1+\alpha _1 {\overline I}} = 0,\label{ss1} 
 \end{equation}
 \begin{equation}
 -\mu _1 {\overline M}_I+\frac{\beta _1 {\overline M}_S {\overline I}}{1+\alpha _1 {\overline I}} = 0, \label{ss2} 
 \end{equation}
 \begin{equation}
k(1-{\overline A})\left({\overline M}_S+ {\overline M}_I\right)-\mu _2{\overline A}-\gamma {\overline A} = 0,\label{ss3} 
\end{equation}
\begin{equation}
 \mu _H-\mu _H {\overline H}-\frac{\beta _2 {\overline H} {\overline M}_I}{1+\alpha _2{\overline M}_I} = 0, \label{ss4} 
 \end{equation}
 \begin{equation}
 \frac{\beta _2 {\overline H} {\overline M}_I}{1+\alpha _2 {\overline M}_I}-\sigma {\overline I}-\mu _H {\overline I} = 0.\label{ss5}
 \end{equation}
From \eqref{ss1} and \eqref{ss2} we obtain
\begin{equation}
\label{ss6}
 {\overline M}_S+ {\overline M}_I=\frac{\gamma {\overline A}}{k\mu_1 +\gamma {\overline A}}.
\end{equation}
From \eqref{ss3} it follows
\begin{equation}
{\overline A}=\frac{k({\overline M}_S+ {\overline M}_I)}{\mu_2 +\gamma +k({\overline M}_S+ {\overline M}_I)}. \label{ss7}
\end{equation}
Adding \eqref{ss4} and \eqref{ss5} we obtain
\begin{equation}
\label{ss8}
{\overline H}=1-\frac{\sigma +\mu_H}{\mu_H} {\overline I}.
\end{equation}
Now, in order to get the disease--free states, we distinguish two cases:\\
\emph{(i)} If ${\overline M}_S={\overline M}_I={\overline A}={\overline I}=0$ we obtain from \eqref{ss8} ${\overline H}=1$.\\
\emph{(ii)} If ${\overline M}_I={\overline I}=0$ and ${\overline M}_S\neq 0$, ${\overline A}\neq 0$, we obtain from \eqref{ss8} ${\overline H}=1$, and substituting \eqref{ss6} into \eqref{ss7} we obtain 
$$
\frac{k\gamma}{(\mu _2+\gamma )(k\mu _1+\gamma {\overline A})+k\gamma {\overline A}}=1,
$$
from which we get
$$
{\overline A}=\frac{k(\gamma -\mu _1(\mu_2 +\gamma))}{\gamma (\mu _2+\gamma +k)}.
$$
Taking into account of (\ref{Q0}), we have
$$
{\overline A}=\frac{k(1-Q_0^{-1})}{\mu _2+\gamma +k},
$$
and substituting ${\overline A}$ in \eqref{ss6} we obtain 
$$\displaystyle {\overline M}_S=\frac{\gamma (1-Q_0^{-1})}{\mu _1k+\gamma}.$$
Therefore, the disease--free equilibria are given by (\ref{E0}) and (\ref{E1}).

\section{Derivation of condition (\ref{condLAS})}
\label{Derivation}

Here we prove that the inequality (\ref{SC}) may be written as (\ref{condLAS}).\\
Let us begin by observing that from (\ref{ss4}) and (\ref{ss5}) we get
$$
1-H^*=\frac{(\sigma+\mu_H)}{\mu_H} I^*,
$$
and hence
$$
M_I^*=\frac{(\sigma+\mu_H)I^*}{\beta_2-\frac{\beta_2(\sigma+\mu_H)}{\mu_H}I^*-\alpha_2(\sigma+\mu_H)I^*},
$$
which can be written
\begin{equation}
\label{MI2}
M_I^*=\frac{(\sigma+\mu_H)I^*}{\beta_2-\frac{(\beta_2+\alpha_2)(\sigma+\mu_H)}{\mu_H}I^*}.
\end{equation}
On the other hand, from (\ref{ss5}) we have:
$$I^*=K_1 \left(R_0^2-1\right)
$$
where $K_1$ is given by (\ref{K1}). Substituting (\ref{MI2}) in (\ref{SC}) we have:
$$
\frac{\beta_1\beta_2}{\mu_1}\frac{(I^*)^2}{\beta_2-\frac{(\beta_2+\alpha_2)(\sigma+\mu_H)}{\mu_H}I^*}>1.
$$
Taking into account of (\ref{Istar}), this can be written
$$
K_0 K_1^2 \left(R_0^2-1\right)^2 > \beta_2-K_2 K_1 \left(R_0^2-1\right),
$$
where $K_0$, $K_1$ and $K_2$ are given by (\ref{K0}), (\ref{K1}) and (\ref{K2}) respectively. Therefore, we have:
$$
K_0 K_1^2 x^2 +K_2 K_1 x - \beta_2 > 0,
$$
where $x=\left(R_0^2-1\right)$.
This inequality is satisfied in the exterior of the interval $\left(x_1,x_2\right)$
where
$$
x_{1,2}=\frac{-K_1 K_2 \mp \sqrt{K_1^2K_2^2+4K_0K^2_1\beta_2}}{2K_0K_1^2},
$$
that is, for:
$$
R_0^2<1+x_1,
$$
which cannot be considered because $x_1$ is negative and here we are assuming $R^2_0>1$, because this guarantee the existence of the endemic equilibrium, or
$$
R_0^2>1+x_2,
$$
and the inequality (\ref{condLAS}) follows. 

\section{Proof of Theorem \ref{Forward}}
\label{proofbif}

The Jacobian matrix of model (\ref{ec1}) evaluated at $E_1$ for $\beta_1=\beta_1^*$ is
$$
J(E_1,\beta_1^*)=  \left(\begin {array}{ccccc}
 -\frac{\gamma}{k}a^*-\mu_1 & -\frac{\gamma}{k} a^* & \frac{\gamma}{k}(1-m^*) & 0 & -\beta_1m^*g'_1(0) \\
  0 & -\mu_1 & 0 & 0 & \beta_1m^*g'_1(0) \\
k(1-a^*) & k(1-a^*) & -km^*-\mu_2-\gamma & 0 & 0 \\
0 & -\beta_2g'_2(0) & 0 & -\mu_H & 0 \\
0 & \beta_2g'_2(0) & 0 & 0 & -\sigma-\mu_H 
\end {array} \right),
$$
It admits a simple zero eigenvalue and the other eigenvalues are real and negative. Hence, when $\beta_1=\beta_1^*$ (or, equivalently, when $R_{0}=1$),  the disease-free equilibrium $E_1$ is a nonhyperbolic equilibrium.\\
Note that
$$
g'_1(0)=1;\;\;\;\;g'_2(0)=1.
$$
Denote by  ${\bf v}=(v_{1},v_{2},v_{3})$, and ${\bf
w}=(w_{1},w_{2},w_{3})^{T}$, a left and a right eigenvector
associated with the zero eigenvalue; that is $J(E_1,\beta_1^*){\bf w}={\bf 0}$, and ${\bf v}J(E_1,\beta_1^*)={\bf 0}$. Require also that ${\bf v} \cdot {\bf w}=1$. It can be easily checked that 
$$
{\bf v}=\left(0,\;\frac{\beta_2}{\alpha_2\left(\sigma+\mu_H+\mu_1\right)},\; 0,\; 0,\; \frac{\mu_1}{\sigma+\mu_H+\mu_1} \right), 
$$
and
$$
{\bf w}=
\left( -\frac{\alpha_2\left(\sigma+\mu_H\right)}{\beta_2},\; \frac{\alpha_2\left(\sigma+\mu_H\right)}{\beta_2},\; 0,\; - \frac{\left(\sigma+\mu_H\right)}{\mu_H},\; 1
\right)^{T}.
$$
The coefficients $a$ and $b$  may be now explicitly computed. Taking into account of
system (\ref{ec1}) and considering only the nonzero
components of the left eigenvector {\bf v}, it follows that:
$$
\begin{array}{ll}
a = & 2v_2w_1w_5\displaystyle \frac{\partial^{2}f_{2}}{\partial
M_S \partial I} (E_1,\beta_1^*)+v_2w^2_5 \displaystyle \frac{\partial^{2}f_{2}}{\partial I^2} (E_1,\beta_1^*)+\\
&  2v_5w_2w_4\displaystyle \frac{\partial^{2}f_{5}}{\partial
M_I \partial H} (E_1,\beta_1^*)+v_5w^2_2 \displaystyle \frac{\partial^{2}f_{5}}{\partial M_I^2} (E_1,\beta_1^*),
\end{array}
$$
and
$$
\begin{array}{ll}
b =  v_2w_5\displaystyle \frac{\partial^{2}f_{2}}{\partial I
\partial \beta_1} (E_1,\beta_1^*).
\end{array}
$$
It can be checked that:
$$
\begin{array}{l}
\displaystyle \frac{\partial^{2}f_{2}}{\partial M_S \partial I} (E_1,\beta_1^*) = \beta_1^*g'_1(0),\;\;\;\;\;\;\;
\displaystyle \frac{\partial^{2}f_{2}}{\partial I^2} (E_1,\beta_1^*) = \beta_1^* m^* g''_1(0),\\ \\
\displaystyle \frac{\partial^{2}f_{5}}{\partial M_I \partial H} (E_1,\beta_1^*) = \beta_2 g'_2(0),\;\;\;\;\;\;\;
\displaystyle \frac{\partial^{2}f_{5}}{\partial M_I^2} (E_1,\beta_1^*) = \beta_2 g''_2(0),\\ \\
\displaystyle \frac{\partial^{2}f_{2}}{\partial I \partial \beta_1} (E_1,\beta_1^*)= m^* g'_1(0), \\ \\
\end{array}
$$
where $g''_1(0)=-2\alpha_1 $, and $g''_2(0)=-2 \alpha_2$.

Now, by checking the signs term-by-term, it is easy to conclude that $a<0$ and $b>0$. Therefore, system
(\ref{ec1}) exhibits  forward bifurcation at $E_1$ and  $R_0=1$. $\#$



\begin{thebibliography}{99}
\bibitem{adetal} Adongo D, Fister K.R., Gaff H., Hartley D, Optimal control applied to rift valley fever,    
Nat. Res. Modeling, 26, 385--402 (2013)
\bibitem{anma} Anderson R. M., May R. M., Infectious Diseases in Humans: Dynamics and Control. Oxford University Press, Oxford, 1991
\bibitem{anca} Anita S., Capasso V.,  Stabilization of a reaction--diffusion system modelling malaria transmission, 
Disc. Cont. Dyn. Sys. B, 17, 1673--1684 (2012)
\bibitem{beetal} Benedict M.Q., Levine R.S., Hawley W.A., Lounibos L.P., Spread of the tiger: global risk of invasion by the mosquito \emph{Aedes albopictus}., Vector Borne Zoonotic Dis., 7, 76--85 (2007)
\bibitem{bbcruz2} Buonomo B., Vargas De--Le\'on C., Stability and bifurcation analysis of a vector-bias model of malaria transmission. Math. Biosci.,  242, 59--67 (2013) 
\bibitem{cagu} Cai L., Guo S., Li X. Z., Ghosh M., Global dynamics of a dengue epidemic mathematical model, Chaos, Soliton. Fract., 42, 2297–-2304 (2009)
\bibitem{cali} Cai L. M., Li X. Z., Global analysis of a vector-host epidemic model with nonlinear incidences. Appl. Math. Comput., 217, 3531--3541 (2010) 
\bibitem{cap} Capasso V., Mathematical Structures of Epidemic Systems. Lecture Notes in Biomath., 97. Springer-Verlag, Berlin, 1993
\bibitem{cagrse} Capasso V., Grosso E., Serio G., I modelli matematici nella indagine epidemiologica. Applicazione all'epidemia di colera verificatasi in Bari nel 1973 (italian), Annali sclavo, 19,  193--208 (1977) 
\bibitem{case} Capasso V., Serio G., A generalization of the Kermack--Mc Kendrick deterministic epidemic model, Math. Biosci., 42  41--61 (1978) 
\bibitem{ccso} Castillo--Chavez C., Song B., Dynamical models of tuberculosis and their applications. Math. Biosci. Engin., 1, 361--404 (2004)
\bibitem{chitetal} Chitnis N., Hyman J. M., Cushing J. M., Determining important parameters in the spread of malaria
through the sensitivity analysis of a mathematical model. Bull. Math. Biol., 70, 1272--1296 (2008)
\bibitem{chetal} Chowell G., Castillo--Chavez C., Fenimore P. W., Kribs--Zaleta C. M., Arriola L., Hyman J. M., Model Parameters and Outbreak Control for SARS, Emerg. Infect. Dis., 10, 1258-–1263 (2004)
\bibitem{dihe} Diekmann O., Heesterbeek J. A. P., Mathematical Epidemiology of Infectious Diseases. Model building, analysis and interpretation. John Wiley \& Sons, Chichester, 2000
\bibitem{do} d'Onofrio A., Vaccination policies and nonlinear force of infection: generalization of an observation by Alexander and Moghadas (2004), Appl. Math. Comput.,  168, 613--622 (2005)
\bibitem{dudu} Dumont Y., Dufourd C., Spatio--temporal modeling of mosquito distribution, AIP Conference Proceedings
Vol.1404, 162--167 (2011)
\bibitem{duhucc} Dushoff J., Huang W., Castillo--Chavez C., Backward bifurcations and catastrophe in simple models of fatal diseases, J. Math. Biol., 36, 227--248, 1998.
\bibitem{esma} Esteva L., Matias M., A model for vector transmitted diseases with saturation incidence.  J. Biol. Sys., 9, 235--245 (2001) 
\bibitem{esva1} Esteva L., Vargas C., Analysis of a dengue transmission model, Math. Biosci., 150, 131--151 (1998).
\bibitem{esva2} Esteva L., Vargas C., Influence of vertical and mechanical transmission on the dynamic of dengue disease, Math. Biosci., 167, 51--64 (2000).
\bibitem{guho} Guckenheimer, J., Holmes, P., Nonlinear Oscillations, Dynamical Systems and Bifurcations of Vector Fields, Springer-Verlag, Berlin, 1983.
\bibitem{he} Hethcote H. W. , The mathematics of infectious diseases, SIAM Rev., 42, 599--653 (2000)
\bibitem {kero} Keeling M. J., Rohani P., Modeling Infectious Diseases in Humans and Animals. Princeton University Press, Princeton, 2008
\bibitem{luetal}  Lutambi A. M., Penny M. A., Smith T., Chitnis N., Mathematical modelling of mosquito dispersal in a heterogeneous environment, Math. Biosci., 241, 198--216 (2013)
\bibitem{maya} Maidana N. A., Yang H. M., Describing the geographic spread of dengue disease by traveling waves. Math Biosci., 215, 64--77 (2008)
\bibitem{maetal} Marino S., Hogue I. B., Ray C. J., Kirschner D. E., A methodology for performing global uncertainty and sensitivity analysis in systems biology, J. Theor. Biol., 254, 178--196 (2008)
\bibitem{medebr} Mecoli M., De Angelis V., Brailsford S. C., Using system dynamics to evaluate control strategies for mosquito-borne diseases spread by human travel, Comput. Operat. Res. 40, 2219--2228 (2013)
\bibitem{glgo} Moreno Glasser C., de Castro Gomes A., Infestation of S. Paulo State, Brazil, by \emph{Aedes aegypti} and \emph{Aedes albopictus}. (Portuguese)  Rev. Sa\'ude P\'ublica, 34, 570--577 (2000)
\bibitem{moaaca} Moulay D., Aziz--Alaoui M. A., Cadivel M., The chikungunya disease: Modeling, vector and transmission global dynamics. Math. Biosci., 229, 50--63 (2011)
\bibitem{neetal} Neteler M., Roiz D., Rocchini D., Castellani C., Rizzoli A., Terra and Aqua satellites track tiger mosquito
invasion: modelling the potential distribution of \emph{Aedes albopictus} in north-eastern Italy. Int. J. Health Geogr., 10:49 (2011)
\bibitem{ozetal} Ozair M., Lashari A. A., Jung I. H., Okosun K. O., Stability analysis and optimal control of a vector--borne disease with nonlinear incidence, Discrete Dyn. Nat. Soc., 2012, Article ID 595487 (2012)
\bibitem{poetal} Poletti P., Messeri G., Ajelli M., Vallorani R., Rizzo C., Merler S., Transmission potential of chikungunya virus and control measures: the case of Italy. PLoS One, 3, e18860 (2011)
\bibitem{Rpackage} R Development Core Team: a language and environment for statistical computing. R Foundation for Statistical Computing, Vienna, http://www.r-project.org/foundation
\bibitem{romoto} Rodrigues H. S., Monteiro M. T. T., Torres D. F. M., Vaccination models and optimal control strategies to dengue, Math. Biosci., 247, 1--12 (2014)  
\bibitem{taetal} Takahashi L. T., Maidana N. A., Ferreira W. C. Jr, Pulino P., Yang H. M., Mathematical models for the \emph{Aedes aegypti} dispersal dynamics: travelling waves by wing and wind. Bull. Math. Biol., 67, 509--528 (2005)
\bibitem{to} Tolle M. A., Mosquito-borne diseases. Curr. Probl. Pediatr. Adolesc. Health Care, 39, 97–-140 (2009)
\bibitem {vawa02} van den Driessche P., Watmough J., Reproduction numbers and sub--threshold endemic equilibria for compartmental models of disease transmission, Math. Biosci., 180, 29--48 (2002)
\bibitem{xiru} Xiao D., Ruan, S., Global analysis of an epidemic model with nonmonotone incidence rate.  Math. Biosci., 208,  419--429 (2007)
\bibitem{jiwaxi} Jin Y., Wang W., Xiao S., An SIRS model with a nonlinear incidence rate, Chaos, Soliton. Fract., 34, 1482–-1497 (2007) 
\end{thebibliography}
\end{document}